\documentstyle[12pt]{article}
\def\lapp{{\ \lower 0.6ex \hbox{$\buildrel<\over\sim$}\ }}
\def\gapp{{\ \lower 0.6ex \hbox{$\buildrel>\over\sim$}\ }}
\def\half{{\textstyle{1\over 2}}}
\def\avM{\overline{M}}
\begin{document}
\begin{titlepage}
\vspace*{-1cm}
\begin{flushright}
DTP/95/42   \\
May 1995 \\
\end{flushright}
\vskip 1.cm
\begin{center}
{\Large\bf On the Coulomb Effects on  the \\ [2mm]
 W Line Shape at LEP2}
\vskip 1.cm
{\large  V.A.~Khoze}
\vskip .2cm
{\it Department of Physics, University of Durham \\
Durham DH1 3LE, England }\\
\vskip   .4cm
and
\vskip .4cm
{\large  W.J.~Stirling}
\vskip .2cm
{\it Departments of Physics and Mathematical Sciences, University of Durham \\
Durham DH1 3LE, England }\\
\vskip 1cm
\end{center}
\begin{abstract}
One of the most powerful methods for obtaining a precision measurement
of $M_W$ at LEP2 is from direct reconstruction of the invariant
mass distribution
of $W$ bosons produced in $e^+e^- \to W^+W^-$.
We investigate the effects
on the $W$ line shape, and in particular on the average invariant mass,
 of phase-space and first-order Coulomb
corrections. The latter are shown to have a non-negligible effect
on the reconstructed mass,
inducing shifts of order $-20$~MeV at collision energies
above threshold,   compared
to the Born approximation cross section.
The sign and magnitude of the
effect can be understood in a simple model calculation in which
one of the $W$ bosons is assumed to be stable.
\end{abstract}
\vfill
\end{titlepage}
\newpage

\section{Introduction}
\label{sec:intro}

A precision measurement of the mass of the $W$ boson, $M_W$, is one of
the main objectives of the LEP2 physics programme. There has already
been considerable progress in the precision determination of $M_W$ from
$W\to l\nu$ events at the Tevatron $p \bar p$ collider \cite{CDFMW},
and the level of precision is expected to increase further in the coming
years \cite{CDFFUTURE}. It is important, therefore, that the
potentiality of the $M_W$ measurements at LEP2 should be exploited to
the full.

An obvious requirement for the success of these precise studies is
a high level of reliability of the theoretical  predictions  for the
various experimental observables relating to the different methods
for measuring $M_W$ from the
process $e^+e^-\to W^+W^-$.\footnote{A summary of
these methods can be found,
for example, in Refs.~\cite{AACHEN,DELPHI}.} This in turn requires a
detailed understanding of the physical phenomena which describe the
production and decay of $W$ bosons at LEP2, in particular of the effects
which arise from the large $W$ boson decay width $\Gamma_W$ ($
\simeq 2.1$~GeV) \cite{KHOZE}. The instability of the $W$ bosons can,
in principle, strongly modify the standard `stable $W$' results. For
example, an important role can be played by the QED and QCD radiative
interferences (both virtual and real)  which interconnect the production
and decay stages.  Particular attention should be paid here  to the
virtual contributions corresponding to the `charged' particle poles for
which, in contrast to the photon and gluon poles,  there is no
cancellation from real emissions \cite{FKM1,FKM2} (see also
\cite{MELYAK1,MELYAK2}).     The level of suppression of the
width-induced radiative effects depends on the `degree of inclusiveness'
of the process. Thus, in the case of the total inclusive cross section
for $e^+e^- \to W^+W^- \to 4f$, the final-state interaction effects are
cancelled up to terms of relative order  $\alpha \Gamma_W/M_W$
or $\alpha_s^2 \Gamma_W/M_W$. The only exception is the contribution
arising from the Coulomb interaction between the slowly moving  $W$
bosons \cite{FKM3,FKMC} (see also \cite{BARDIN}). If the $W$ bosons
were {\it stable} particles,
the effect of the Coulomb interaction on the total cross section has
been known for a long time \cite{SOMMERFELD}.
The modifications to the Coulomb corrections which arise from the
{\it instability} of the $W$ bosons are particularly  significant near
the $W^+W^-$ production threshold, $\sqrt{s} \approx 2 M_W$, but
become negligible  for collision energies which satisfy
$\sqrt s  - 2 M_W \gg \Gamma_W$. As explained in detail in
Ref.~\cite{FKMC}, the $W$-boson virtuality drastically changes the
on-shell value of the Coulomb correction even at
$ \sqrt s - 2 M_W
\gg \Gamma_W$, but after integration over the invariant masses of the
two $W$ bosons the `stable-$W$' result is restored far above the
threshold region.

The electroweak radiative corrections to the $e^+e^- \to W^+W^-$ total
cross section  for the production of
 stable $W$ bosons are now known with high accuracy
\cite{BEENDENN}. The instability effects are well under control
throughout the energy range accessible at LEP2 ($\sqrt s \lapp
200$~GeV).
This  includes the so-called `colour-reconnection effects'
\cite{GUSTAFSON,TSVAK,GUSHAK}  --- non-perturbative hadronization dynamics
which may affect the $W^+W^-\to 4q$ decay channel and which have recently
attracted much attention.

Two different methods have been proposed for a precision measurement
of $M_W$ at LEP2. The first uses the method of direct
kinematic reconstruction of $M_W$ from the decay channels
\begin{eqnarray}
\label{fourq}
W^+W^-        &\to& q \bar q q \bar q , \\
\label{semilep}
W^+W^-        &\to& q \bar q  l \nu .
\end{eqnarray}
The second method uses the strong dependence of the total
$W^+W^-$ cross section on $M_W$ near threshold to translate
a measurement of $\sigma_{WW}$ close to $\sqrt{s} = 2 M_W$
into a value of $M_W$.
 Both methods have been,
and are currently being, studied in great detail, see for example
Refs.~\cite{AACHEN,DELPHI,WJS} and references therein.
Although both have their advantages and disadvantages, the
direct reconstruction method is currently believed to provide the
better precision on $M_W$. Since it naturally requires
a higher collision energy to maximise the event rate, it does not
conflict with the requirements of new particle/physics searches.

However, the direct reconstruction method is not without problems. For
example, to construct the two $W$'s from the $q \bar q q \bar q$
final state in  (\ref{fourq}) one must in principle attribute all
observed hadrons to the `correct' parent $W$, a procedure which is
certainly affected by relatively unknown QCD interconnection
corrections \cite{TSVAK}.
Since a complete description of these effects is not possible at
present, one has to rely on model predictions rather than on exact
calculations; for details see Refs.~\cite{TSVAK,GUSHAK}.
Fortunately, the contribution of the interconnection effects  to the
systematic error on $M_W$ is not expected to significantly exceed the
overall systematic error (not including interconnection effects),
which is currently estimated at $O(30-40)$~MeV.
It is an open question whether Bose-Einstein effects might induce
a further uncertainty in the mass determination \cite{TSVAK,LONNBLAD}.
 Such problems do not
of course arise  for the semi-leptonic channel (\ref{semilep}),
but there the event rate is smaller and an unobserved neutrino is
present.

The direct reconstruction method is based on measurements of
semi-inclusive characteristics of the final state --- $W$ boson momenta,
jet--jet invariant masses, opening angle between jets, etc. All such
quantities could well be much more sensitive to instability effects than
the total cross section. For example, as could be anticipated from
 Ref.~\cite{FKMC}
(see also below), the modification of the QED Coulomb interaction could
induce a systematic shift in the $W$ mass of $O(\alpha \pi \Gamma_W)$,
which is of the same order as the target precision. Clearly such effects
should be carefully calculated and taken into account in the extraction
of the $W$ mass from measured distributions. Note that since Coulomb
forces are responsible for the QED interaction between the separate
hadronic or leptonic final states of each $W$, their effects could be
regarded as an example of `QED interconnection effects'. These effects
are in fact quite universal and do not depend on the particular $W$
decay channels.

In the threshold region the Coulomb contribution dominates the instability
effects, and the Coulomb terms can be uniquely separated from the remaining
electroweak corrections. At higher energies, the width-induced
modifications of the differential distributions caused by other radiative
mechanisms (for example, intermediate--final and final--final state
radiative interferences \cite{FKM1,MELYAK2} ) may become
just as important.\footnote{Recall that in the relativistic region the
Coulomb term is neither uniquely defined nor separately gauge invariant.}
Moreover, it is argued in Ref.~\cite{MELYAK2} that in the relativistic
domain a cancellation may take place between the different sources  of
instability effects and
that, as a result, the stable $W^+W^-$ result may be restored.
In the extreme relativistic limit, $(1-\beta) \ll 1$, where $\beta$ is the $W$
velocity, such a cancellation appears quite
natural.\footnote{It has its origins in the conservation of `charged' currents.
Another example of the vanishing of off-shell effects at very high energies
was discussed in Ref.~\cite{DKT} (see also \cite{KOS}),
where the gluon radiation pattern
corresponding to top quark production and decay was discussed.
For the most probable kinematic configuration (quasi-collinear $b$ and
$t$), the width dependence disappears completely when the effects
of emission at the production and decay stages are added coherently.
The same behaviour is expected for QED radiation off fast-moving $W$
bosons \cite{DKOS}.}
However in the intermediate region $\beta < 1$, which is relevant for
the LEP2 energy range,
 the situation is less clear in our opinion,
  and therefore needs further detailed study.
Since $\sqrt{s} - M_W \ll M_W$ at LEP2, the non-relativistic
Coulomb formulae are likely to provide a reasonable qualitative
guide to the size of the width-induced effects.

In this  study we analyse the effect of the $W$ boson instability, as embodied
in the first-order Coulomb formulae of Refs.~\cite{FKM3,FKMC},
on the invariant mass distribution of the decay
products. As far as we are aware, the necessity to take
Coulomb-induced distortion effects in the $W$ mass or momentum
distribution into account was first pointed out in Ref.~\cite{TSVAK}.
In order to expose the direct effect of this `QED interconnection',
we make several simplifying assumptions in our analysis:
(i) the effects of initial state radiation are not included,
although it would be straightforward to take them into account using the
standard techniques, (ii) we assume that the  $W^+W^-$ final state can be fully
reconstructed, and (iii) we neglect possible QCD interconnection effects.
Note that our results also apply directly to the process
$\gamma\gamma\to W^+W^-$.

The paper is organized as follows. In Section~\ref{sec:model}
 we present  a simple
model calculation  in which one of the $W$ bosons is assumed to be stable.
Above threshold this is expected  to provide a reasonable
qualitative understanding.
Using this model, we study the effect on the invariant mass distribution
of the decaying $W$ of phase-space and first-order Coulomb effects. In
Section~\ref{sec:quant} we study numerically the realistic case when
both $W$ bosons are off-mass-shell. Predictions for the shift in the average
mass with and without Coulomb corrections are presented. Finally,
Section~\ref{sec:conc} contains our conclusions.

\section{A model analysis}
\label{sec:model}
To elucidate the physical origin of the distortion of the  $W$ decay mass
distribution induced
by the Coulomb interaction, it is instructive to consider first a simplified
model in which one of the $W$ bosons is assumed to be stable and the other has
the standard $W \to f \bar f$ decay modes with decay width
$\Gamma_W$\footnote{In fact precisely this situation applies to the
production of a charged Higgs boson with $M_H \simeq M_W$ in
$e^+e^- \to Z^* \to W^\pm H^\mp$, since in most models $\Gamma_H
\ll \Gamma_W$. The observation of such a process would be a signature
of an exotic Higgs sector, see for example Ref.~\cite{IUK}.}
(see also \cite{MELYAK2}).

In the non-relativistic region the differential distribution of the
invariant mass-squared $s_1$ of the unstable $W$ can be written as
\begin{equation}
\label{dsig}
{d \sigma \over d s_1} \approx \rho(s_1) \sigma_0(s,s_1,M_W^2)
\left[ \left( 1 + {\alpha\over \pi} \delta_H \right) +
{\alpha\over 2 \beta} \delta_C \right],
\end{equation}
where $\sigma_0(s,s_1,s_2)$ is the  $e^+e^-\to W^+W^-$
off-shell  Born cross section \cite{MUTA} at centre-of-mass energy
$\sqrt{s}$, and $\rho(s_i)$ is the Breit-Wigner factor\footnote{We omit
here a trivial overall branching ratio factor for the particular
$f \bar f$ final state under consideration.}
\begin{equation}
\rho(s_i) = {1\over \pi}\; {\sqrt{s_i}\; \Gamma_W(s_i)\over
(s_i-M_W^2)^2 + s_i \Gamma_W^2(s_i) }  ,
\label{rho}
\end{equation}
with $\Gamma_W(s_i) = \sqrt{s_i}\; \Gamma_W /M_W$.
The coefficient $\delta_H$ is the `hard' first-order radiative
correction, $(\alpha/ 2 \beta) \delta_C$ is the first-order Coulomb
contribution (see Refs.~\cite{FKM3,FKMC}), and $\beta$ is the velocity
of the $W$ bosons in the centre-of-mass frame.  In the non-relativistic
approximation $\beta = 2 p / \sqrt{s}$ where
\begin{equation}
p^2 \approx (\sqrt{s} - \sqrt{s_1} - M_W) M_W .
\end{equation}
In our model calculation it will be sufficient to work in the
non-relativistic approximation. It is then straightforward to show that
in this case $\delta_C$ is given by the same formulae as in the
realistic case of two unstable $W$ bosons \cite{FKM3,FKMC},
with $\Gamma_W$ replaced by $\Gamma_W/2$. Thus
\begin{equation}
\delta_C =
\pi -2\; \mbox{arctan}\left({\vert\kappa\vert^2 -p^2  \over
 2 p\; {\rm Re}(\kappa ) }  \right) ,
\label{omega}
\end{equation}
with
\begin{equation}
\kappa = \sqrt{-M_W(E+ \half i \Gamma_W)}.
\label{kappa}
\end{equation}
Here $E$ is the non-relativistic energy of the $W$ bosons,
\begin{equation}
E \approx \sqrt{s} - 2 M_W.
\end{equation}
Recall that as a direct consequence of the dominance of  $S$-wave
$W^+W^-$ production at $\beta \ll 1$,\footnote{Throughout this paper
we assume that the $\nu$--exchange contribution, which dominates the
$e^+e^-\to W^+W^-$ threshold cross section, is not suppressed by a
particular choice of $e^\pm $ beam polarizations. The mass distortion
effects discussed here are strictly only valid for {\it unpolarized}
scattering.}
\begin{equation}
\label{swave}
\sigma_0 \simeq \mbox{const.}\;\times \beta + O(\beta^3) .
\end{equation}
It is worth mentioning that the higher-order terms in the
expansion (\ref{swave}) lead to a net negative correction to the
leading $\sim \beta$ behaviour, see for example Fig.~1 of
Ref.~\cite{WJS}, which originates in the high-energy SU(2)$\times$U(1) gauge
invariance cancellation.

For $E \gg \Gamma_W$ and in the dominant (`peak') region
specified by $\vert s_1 - m_W^2 \vert \lapp  M_W \Gamma_W$, one finds
\begin{equation}
\label{peakapprox}
\delta_C \simeq
\pi -2\; \mbox{arctan}\left({ s_1  -M_W^2  \over
  M_W  \Gamma_W  }  \right) .
\end{equation}
Eq.~(\ref{peakapprox}) reveals the strong dependence of
the coefficient $\delta_C$ on the $W$ boson virtuality $s_1$.
This follows from the general nature of the Coulomb forces between
unstable heavy particles \cite{FKMC}. Thus in the large invariant mass
tail ($s_1  > M_W^2$), $\delta_C$ is strongly suppressed, while in the
small invariant mass tail ($s_1 < M_W^2$), $\delta_C \approx 2 \pi$,
i.e. twice the first-order on-mass-shell value.
After integration over $s_1$, the arctan modification of $\delta_C$
averages to zero and the stable $W$ result obtains.
Note that while for the total $e^+e^- \to W^+W^-$
cross section the stable $W$ result for the Coulomb correction is only
strongly modified by instability effects in the narrow energy region
$E  \lapp \Gamma_W$ (i.e. close to threshold), for
 the invariant mass distribution  the arctan modification of $\delta_C$
 is essentially  independent of energy for $E \gapp \Gamma_W$.
 However, far above threshold additional energy dependence {\it will}
 appear due to the screening role of the other QED final-state interaction
 mechanisms.

For purposes of illustration, it is convenient to rewrite the cross
section formula (\ref{dsig}) for $\Gamma_W \ll E \ll M_W$ in the dominant
$s_1$ (peak) region in terms of the dimensionless  variable $x$ where
\begin{equation}
\label{xdef}
x= { s_1  -M_W^2  \over M_W  \Gamma_W  } \sim O(1) .
\end{equation}
The differential cross section in Eq.~(\ref{dsig}) then becomes
\begin{equation}
\label{dsigx}
{d \sigma \over d x} \approx  { \sigma_0 \over \pi}\; {1\over 1 + x^2} \;
\left[ \left( 1 + {\alpha\over \pi} \delta_H \right) +
{\alpha\over  \beta} \left( {\pi\over 2}-\mbox{arctan}\/x \right) \right],
\end{equation}
with
\begin{equation}
\beta= \sqrt{ { E - \half x \Gamma_W   \over M_W }  }  .
\end{equation}
The invariant mass distribution $d\sigma / d x$ deviates from the
Breit-Wigner $(1+x^2)^{-1}$ form, corresponding to an
individual $W$ decay,  because of (i) the strong dependence of
the threshold Born cross section $\sigma_0$ on the $W$ momentum,
and (ii) the characteristic behaviour of the Coulomb term discussed
above. In particular, when $\beta \ll 1$, which corresponds to
$x \sim 2 E /\Gamma_W$, the invariant mass distribution is strongly
suppressed by phase space effects, Eq.~(\ref{swave}). These lead
to a decrease in the average value of  the invariant mass, $M = \langle
\sqrt{s_1} \rangle$, in the threshold region by\footnote{The subscript
`$B$' denotes a shift in the average mass  due to the Born cross section
behaviour.}
\begin{eqnarray}
\Delta M_B &=& O\left( {\Gamma_W^2 \over E } \right) \ \mbox{at}\
 E \gg \Gamma_W  , \nonumber  \\
\Delta M_B &=& O\left( \Gamma_W\right) \ \mbox{at}\
 \vert E\vert  \le \Gamma_W  , \nonumber  \\
\Delta M_B &=& O\left( \vert E\vert \right) \ \mbox{at}\
-E  \gg \Gamma_W  .
\end{eqnarray}
With increasing $\beta$ the negative higher-order (in $\beta^2$) terms
in the expansion (\ref{swave}) become more and more important and, as a result,
$ \Delta M_B $ changes sign (see Section~\ref{sec:quant} below).

The characteristic dependence on the $W$ virtuality of the Coulomb correction
for $E \gg \Gamma_W$ always causes a decrease in the average mass compared
to the Born prediction:
\begin{equation}
\Delta M_C = O(\pi \alpha \Gamma_W).
\end{equation}
Note that this mainly arises from the Coulomb-induced asymmetry
in the tails of the distribution, as discussed above.
The shift in the actual position of the peak  is numerically
rather small.

It is important to emphasize the difference between the predictions
of Eqs.(\ref{omega},\ref{peakapprox}) and the `on-mass-shell Coulomb'
 correction, $\alpha\pi/(2\beta)$.
The latter would induce a shift in the $s_1$ distribution
towards larger values, throughout the threshold region.

Note that there appears to be a  range  of collider
energies around $\sqrt{s} \approx 190$~GeV where the phase space and Coulomb
induced distortions are of the same order in their effect on
the average mass (see Figs.~3  and 4 below).
However at this energy, which is of practical importance for LEP2, other
mechanisms (e.g. intermediate--final and final--final radiative
interference involving the decay products of the two
$W$ bosons) have to be taken into account.

Finally,  higher-order  ($O(\alpha^n), \; n \geq 2$)
Coulomb effects could  be numerically more important for the invariant mass
distribution than for the total cross section.
In principle, it is straightforward to take these into account
using the general formalism presented in Ref.~\cite{FKMC} (see also
\cite{FADKHO,FKK}).

In summary, we have investigated the qualitative effects
on the invariant mass distribution of phase-space and Coulomb
corrections using a simple model in which only one $W$ boson
is off-mass-shell. In the following section, we shall study numerically
the more realistic case of {\it both} $W$ bosons being off-shell.
As we shall see, the conclusions obtained from our model are
unchanged by the more complete analysis.

\section{Invariant mass distributions
in $e^+e^- \to W^+W^-$: quantitative discussion}
\label{sec:quant}

In the realistic  case when {\it both}
$W$ bosons are off-shell, the $W^+W^-$ cross section can be written
\begin{equation}
\sigma(s) = \int_0^s d s_1 \int_0^{(\sqrt{s}-\sqrt{s_1})^2} ds_2\;
\rho(s_1)\rho(s_2)\; \sigma_0(s,s_1,s_2)
\; \left[  1+ {\alpha\over 2 \beta} \delta_C \right],
\label{sigcoul}
\end{equation}
where $\delta_C$ is again given by Eq.~(\ref{omega})
and $\beta = 2 p /\sqrt{s}$, but now with \cite{FKMC}
\begin{eqnarray}
p^2 & = & {s\over 4} \left[
1-{2s(s_1+s_2)-(s_1-s_2)^2\over s^2}\right], \nonumber \\
\kappa & = &  \sqrt{-M_W(E+i \Gamma_W)},\nonumber \\
E &=& {s - 4M_W^2 \over 4 M_W} .
\label{relforms}
\end{eqnarray}
Note that we have omitted the `hard' radiative corrections and used
the `relativistic' forms for $p^2$ and $E$.
In what follows we will use Eqs.~(\ref{sigcoul},\ref{relforms})
to study (i) the invariant mass distribution $d \sigma/ d s_1$
and (ii) the average invariant mass $\avM$ (which has certain
practical advantages as an estimator of $M_W$ \cite{TSVAK}) defined by
\begin{equation}
\label{average}
\avM = { 1\over \sigma(s)}\; \int_0^s d s_1 \int_0^{(\sqrt{s}-\sqrt{s_1})^2}
ds_2\;
{1\over 2}\left( \sqrt{s_1} + \sqrt{s_2}\right)
\; \rho(s_1)\rho(s_2)\; \sigma_0(s,s_1,s_2)\;
\left[ 1+ {\alpha\over 2 \beta} \delta_C \right].
\end{equation}

Figure~1 shows the normalized distribution $1/\sigma\; d \sigma / d s_1$
(the $W$ `line shape') as a function of $\sqrt{s_1}$,
at three different collider energies, $\sqrt{s} = 165$, $175$
and $185$~GeV.\footnote{We use $M_W = 80.41$~GeV/c$^2$
\cite{CDFMW} and $\Gamma_W = 2.092$~GeV in our numerical
calculations. All other parameters coincide with those used in
Ref.~\cite{WJS} (see Table~1 therein).}
Also shown, for comparison, is the `pure Breit-Wigner' form $\rho(s_1)$
which corresponds, formally, to the $\sqrt{s} \to \infty $ limit.
As anticipated in the previous section, the phase space effects give
a significant distortion to the distribution, especially close
to the $W^+W^-$ threshold. In particular, the distribution is
strongly suppressed for $\sqrt{s_1} \gapp \sqrt{s} - M_W$.

The distributions in Fig.~1 include the
first-order Coulomb correction $\delta_C$. To see the effect of this,
we show in Fig.~2 the ratio
\begin{equation}
f(s,s_1)  = {1\over \rho(s_1)}\ {1\over \sigma} { d \sigma \over  d s_1}
\end{equation}
at $\sqrt{s} = 175$~GeV with and without the Coulomb correction.
Again we confirm the
qualitative behaviour obtained in the model analysis of the previous section:
the Coulomb contribution enhances (suppresses) the small (large) mass tail.
Note, however, that the effect is numerically much less significant
at this energy than the distortion due to phase space effects, which forces
$f$ to be very small for $\sqrt{s_1} \gapp 95$~GeV.

The measurement of the $W$ mass using the
diect reconstruction method at LEP2 involves
fitting  a measured invariant mass distribution,
like that of Fig.~1, by a theoretical distribution (in
practice implemented in a Monte Carlo program) in which
$M_W$ is a free parameter.  In this way, measured quantities
like the position of the peak or the average invariant mass,
both of which are crude measures of $M_W$, are corrected
to the `true' value.
It should be clear from the above discussion that fitting the
measured distribution with a theoretical expression which
does {\it not} include the final-state interaction
effects  will induce an error
in the mass measurement. To quantify this, we focus our attention
on the difference between the average $W$ mass $\avM$ (\ref{average})
defined by the $s_1$ distribution and  the input mass $M_W$,
$\Delta M = \avM - M_W$.
As in the previous section, $\Delta M_B$ denotes the mass shift
using the Born (off-mass-shell) cross section and $\Delta M_C$
denotes the {\it additional} mass shift from
including the $O(\alpha)$ Coulomb correction.

Figure~3 shows $\Delta M_B$ as a function of the collider energy
$\sqrt{s}$. The behaviour can be understood from Fig.~1, and is exactly
as anticipated in Section~\ref{sec:model}. Near and below threshold,
there is a strong phase space suppression for masses $\sqrt{s_1}
> M_W$, and so $\Delta M_B < 0$. Above threshold, the mass difference
grows with increasing collider energy as more and more phase space
for large invariant masses opens up.

A problem with this calculation of $\Delta M$ is that the integrals
over $s_1$ and $s_2$ receive contribution from arbitrarily small
and large invariant masses (subject only to $\sum \sqrt{s_i} \leq
\sqrt{s}$).
In practice, events with very large or very small $f \bar f$ invariant
masses would not be classified as $W$ decay events. In
particular, lower cuts on the $\sqrt{s_i}$ are required to
eliminate non--$W^+W^-$ backgrounds.\footnote{The actual cut value will
in practice depend on the particular final state, collider energy, etc.
We choose two illustrative values for our numerical calculations.}
To make a more realistic calculation, therefore, we impose an
additional cut,
\begin{equation}
\label{range}
\vert \sqrt{s_i} - M_W \vert \leq \delta, \qquad i=1,2.
\end{equation}
Note that this cutting procedure will to some extent complicate
the calculation of the QED and QCD final-state radiative corrections,
because of the reduction of the phase-space for final-state emission.
The mass shifts $\Delta M_B$ for  $\delta = 30$~GeV and 10~GeV are shown as the
dashed and dash-dotted curves respectively
 in Fig.~3. With this additional mass cut
 there is less dependence
on $\sqrt{s}$, since $\Delta M_B \to 0$ as $\delta \to 0$
at fixed $\sqrt{s}$. Note also that  asymptotic values of $\Delta M_B
\to 0.46\ (0.13)$~GeV are approached  for $\delta =
30\ (10)$~GeV, as $\sqrt{s} \to \infty$.\footnote{This is
simply the average value of $\sqrt{s_1}$ weighted by $\rho(s_1)$ over
the range of integration given in Eq.~(\ref{range}).}

Figure~4 shows the additional mass shift $\Delta M_C$
due to the Coulomb correction. Here we see that as long as the mass
cut $\delta$ is not particularly tight, the shift for $\sqrt{s} \gapp
170$~GeV is rather constant at $O(-20)$~MeV. This is consistent
with the model calculation of Section~\ref{sec:model}, which predicted
a constant negative shift of order $\alpha \pi \Gamma_W $ for
$E \gg \Gamma_W$.
As can be derived from  Eqs.~(\ref{sigcoul},\ref{average}),
$\Delta M_C$ changes sign  at lower energies
and attains a maximum  at threshold ($E=0$), where the average
$W$ momentum is lower,
$\langle p \rangle \sim \sqrt{M_W\Gamma_W}$. The actual maximum value
depends on the cut parameter $\delta$. Note once again that
the mass shift decreases at fixed $\sqrt{s}$ as the
invariant mass cut is tightened, i.e. $\Delta M_C \to 0$ as $\delta \to
0$.

\section{Conclusions}
\label{sec:conc}

The success of the precision measurements of the $W$ boson relies
on an accurate theoretical knowledge of the details of the
production and decay mechanisms. The favoured `direct reconstruction'
method of measuring $M_W$ at LEP2 using the hadronic  ($qqqq$) channel
has an important caveat --- the colour reconnection effects induced by the
strong final-state interaction may obscure the separate identities of
the $W$ bosons and thus distort the mass determination \cite{TSVAK}.
At the moment, these effects are not completely curable theoretically
because of the lack of deep understanding  of non-perturbative QCD
dynamics.

However, there are other effects -- originating in purely QED radiative
phenomena -- which, in principle, prevent the final state being treated
as two separate $W$ decays. In this paper we have studied one example
of this, the Coulomb interaction between two unstable $W$ bosons which
induces non-factorizable corrections to the final-state mass
distributions.
Of course there is no reason why all such effects cannot, in principle,
be computed to arbitrary accuracy in QED perturbation theory, and taken
into account in the mass determination. In this paper we have
demonstrated explicitly that their emission could lead
to a $O(20\ {\rm MeV})$
shift in the measured mass.\footnote{It is
also worth mentioning that a similar effect to that described
in this study could also be induced by final-state `new physics'
interactions, for example the exchange between the two $W$ bosons of a
new light scalar with a sufficiently large coupling.} This shift can
only  be  reduced by imposing a rather tight invariant  mass cut, which
selects only those events near the peak if the distribution where the
distortion is minimized.
In particular, we have investigated the effect on the mass distribution
of the QED interconnection effects generated by the first-order Coulomb
corrections in the threshold region at LEP2.  At the highest LEP2
energies, it is likely that we are overestimating the mass  distortion
effect (see for example
 Ref.~\cite{MELYAK2}). In this region, therefore, our results should be
regarded as only a starting point for futher, more detailed studies.
Particularly important in this respect are  the QED
interactions involving the decay products of the two $W$ bosons, which
become essential in the relativistic region.

Finally, we note that similar non-factorizable QED final-state
interaction effects could also be important in precision $M_W$
measurements at the Tevatron $p \bar p$ collider, for example in the
process $q g \to W(\to l \nu) + q$. The distortion would then be
manifest, for example,
in the transverse momentum distribution of the final-state lepton.

\section*{\Large\bf Acknowledgements}

\noindent We are grateful to the UK PPARC for support.
Useful discussions with  Victor Fadin and Torbj\"orn Sj\"ostrand
are acknowledged.
This work was supported in part by the U.S.\ Department of Energy,
under grant DE-FG02-91ER40685 and by the EU Programme
``Human Capital and Mobility'', Network ``Physics at High Energy
Colliders'', contract CHRX-CT93-0319 (DG 12 COMA).
\goodbreak

\section*{Figure Captions}
\begin{itemize}

\item [{[1]}] The distribution $1/\sigma\; d \sigma / d s_1$ in $e^+e^-
\to W^+W^-\to 4f$ production at
 $\sqrt{s} = 165$, $175$ and $185$~GeV. Also shown (dotted line) is the
 asymptotic form, $\rho(s_1)$, given in Eq.~(\ref{rho}).

\item [{[2]}] The ratio of the mass distribution of Fig.~1 to $\rho(s_1)$
at $\sqrt{s} = 175$~GeV, with (solid curve) and without (dashed curve)
the first-order Coulomb correction.

\item [{[3]}] The difference between the average mass $\langle \sqrt{s_1}
\rangle$ and $M_W$, as a function of the collider energy $\sqrt{s}$
(solid curve). Also shown are the mass differences when an additional
cut $\vert \sqrt{s_i} - M_W \vert \leq \delta$ ($\delta = 30,\ 10$~GeV)
is imposed.

\item [{[4]}] The additional mass shift from including the first-order
Coulomb correction, as a function of the collider energy $\sqrt{s}$
(solid curve). Also shown are the mass differences when an additional
cut $\vert \sqrt{s_i} - M_W \vert \leq \delta$ ($\delta = 30,\ 10$~GeV)
is imposed.
\end{itemize}
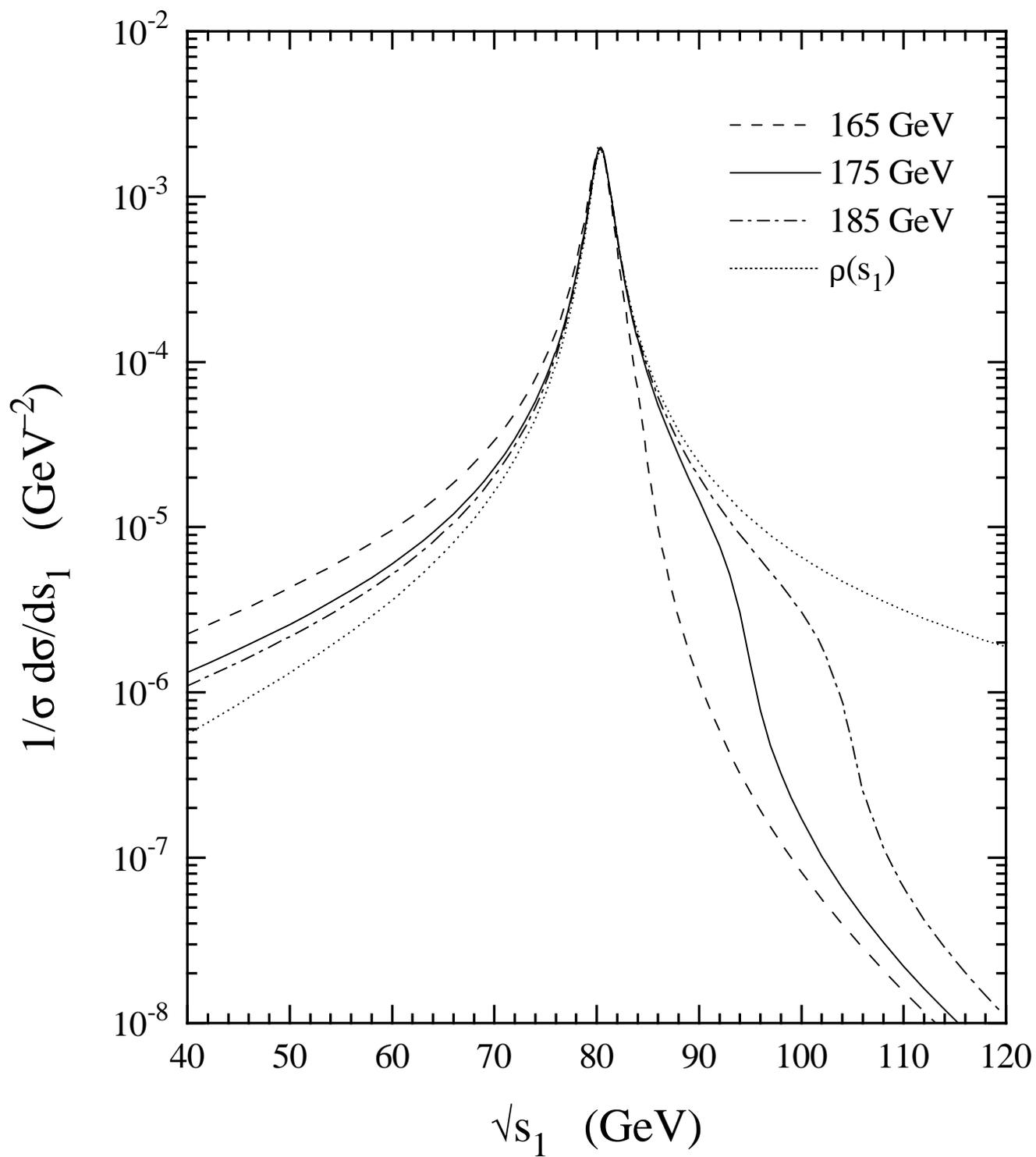
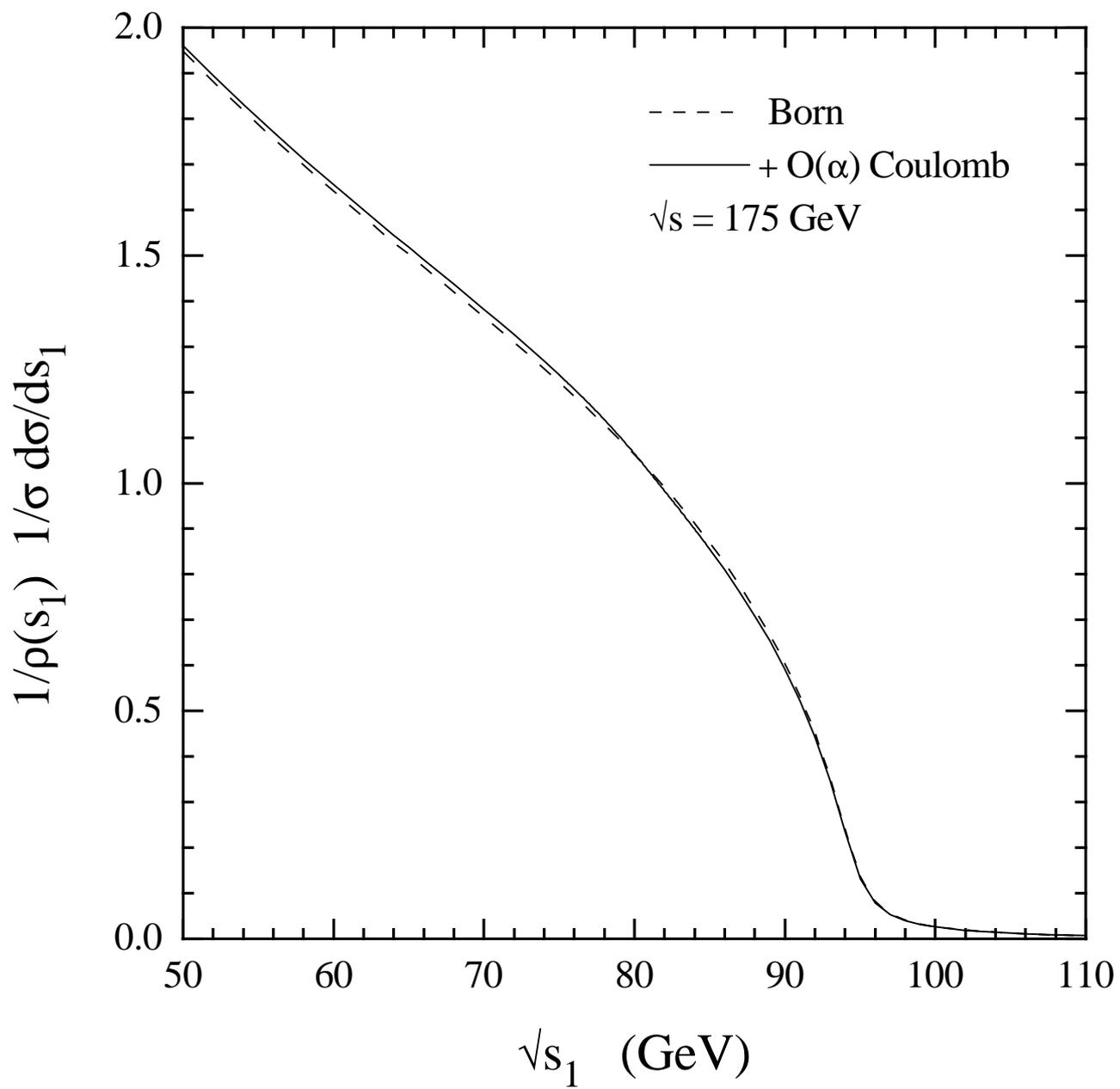
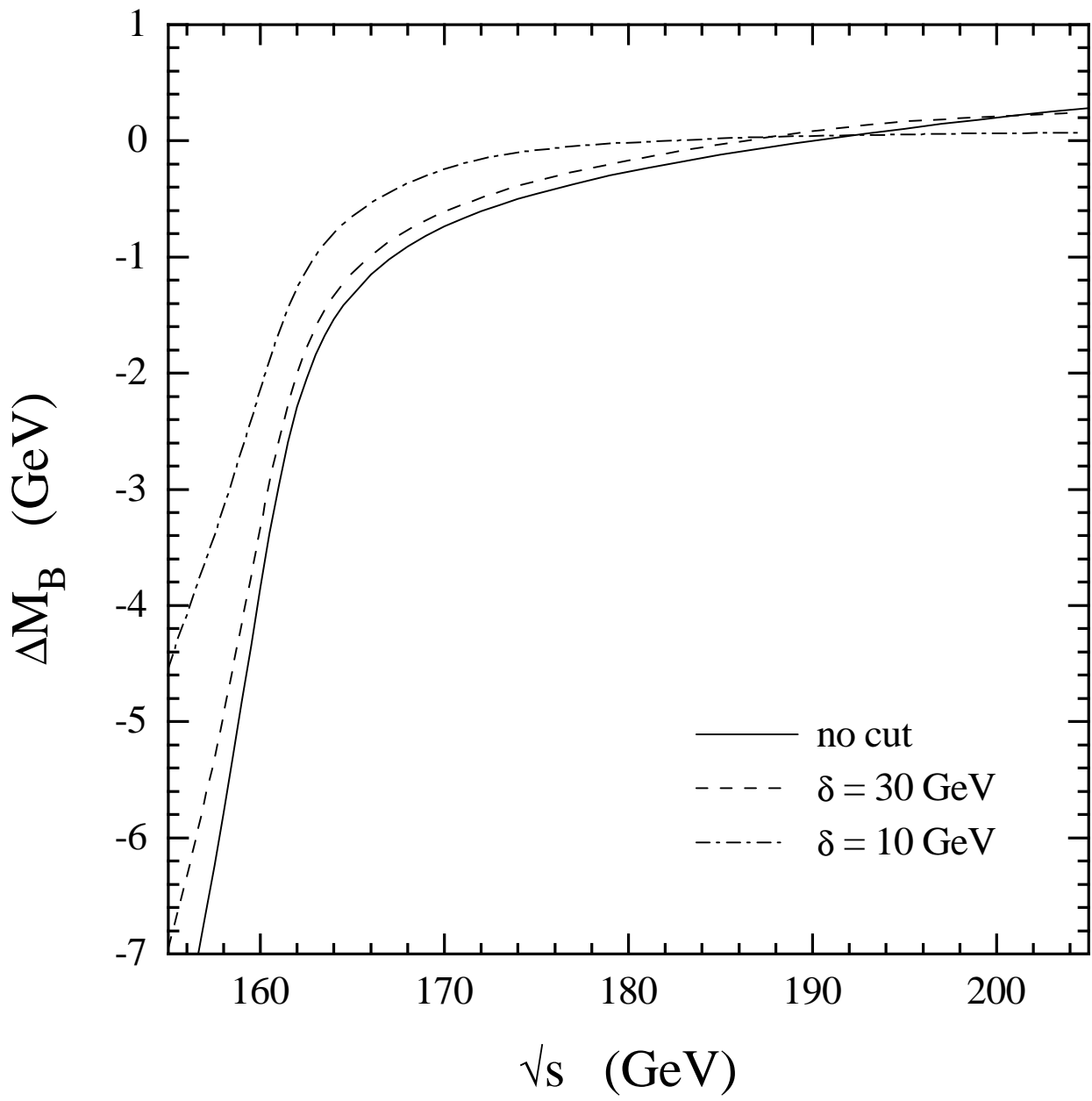
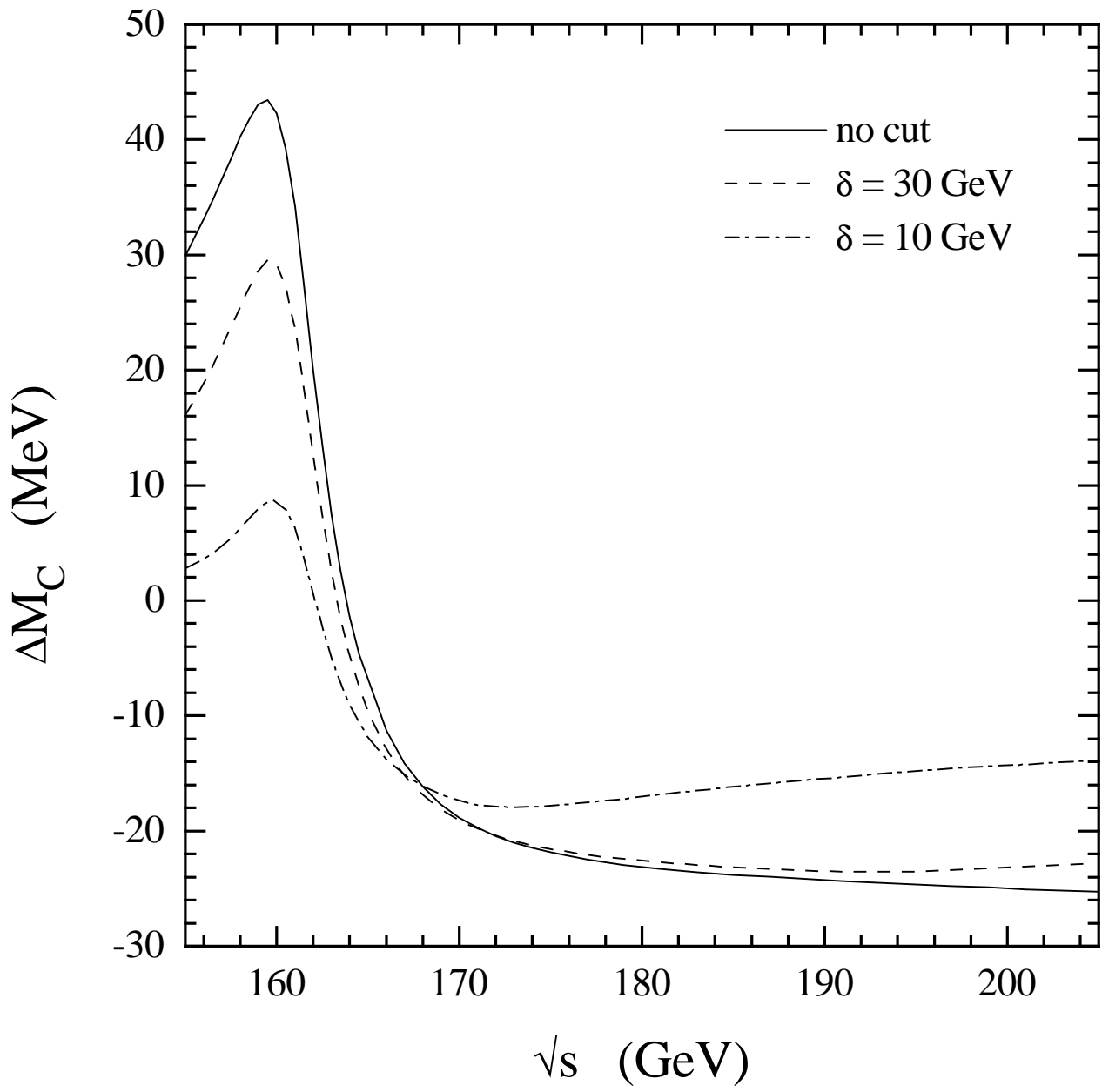
\end{document}